\documentclass[twocolumn,preprintnumbers,amsmath,amssymb]{revtex4}

\usepackage{graphicx}
\usepackage{dcolumn}
\usepackage{bm,epsfig}
\usepackage{wasysym}
\usepackage{amsmath,amssymb}
\usepackage{enumerate}
\usepackage{color}
\usepackage[breaklinks,colorlinks = true,linkcolor = black, urlcolor=black,citecolor = black]{hyperref}

\newcommand{\bs}{\boldsymbol}

\begin{document}


\title{Hall effect anomaly and low-temperature metamagnetism in Kondo compound CeAgBi$_{2}$}

\author{S. M. Thomas}
\thanks{These authors contributed equally to this work.}
\author{P. F. S. Rosa}
\thanks{These authors contributed equally to this work.}
\author{S. B.  Lee}
\author{S.A. Parameswaran}
\author{Z. Fisk}
\author{J. Xia}

\affiliation{
Department of Physics and Astronomy, University of California, Irvine, CA 92697-4574, U.S.A.}
\date{\today}

\begin{abstract}
Heavy fermion (HF) materials  exhibit a rich array of phenomena due to the strong Kondo coupling between their localized moments and itinerant electrons. A central question in their study is to understand the interplay between magnetic order and charge transport, and its role in stabilizing new quantum phases of matter.
Particularly promising in this regard is a family of tetragonal intermetallic compounds Ce{$TX$}$_2$ ($T=$ transition metal, $X=$  pnictogen), that includes a variety of HF compounds showing $T$-linear electronic specific heat $\bf{C_e \sim \gamma T}$, with $\gamma\sim$ 20-500 mJ$\cdot$mol$^{-1}$~K$^{-2}$, reflecting an effective mass enhancement ranging from small to  modest.
Here, we study  the low-temperature field-tuned  phase diagram of high-quality CeAgBi$_2$ using magnetometry and transport measurements.
We find an antiferromagnetic transition at ${T_{N} = 6.4}$~K with weak magnetic anisotropy and the easy axis along the $c$-axis, similar to previous reports (${T_{N} = 6.1}$~K).
This scenario, along with the presence of two anisotropic Ruderman-Kittel-Kasuya-Yosida (RKKY) interactions, leads to a rich field-tuned magnetic phase diagram, consisting of five metamagnetic transitions of both first and second order.
In addition, we unveil an anomalous Hall contribution for fields $H<54$ kOe which is drastically altered when $H$ is tuned through a trio of transitions at 57, 78, and 84~kOe, suggesting that the Fermi surface is reconstructed in a subset of the metamagnetic transitions.
\end{abstract}

\maketitle

In heavy fermion (HF) materials, the Kondo coupling between local moments and itinerant electrons plays a central role in determining magnetic and transport properties, particularly at low temperatures.
Classic examples of the unusual behavior include quantum criticality in YbRh$_{2}$Si$_{2}$~\cite{Custers2003}, unconventional superconductivity in CeCoIn$_{5}$~\cite{CeCoIn5}, and  metamagnetism in CeRu$_{2}$Si$_{2}$~\cite{CeRu2Si2}.
Ce-based HF materials which host such exciting properties often crystallize in tetragonal structures and their ground state is on the border of antiferromagnetism (AFM).

Our focus in this paper is the HF family Ce\emph{TX}$_2$ ($T =$ transition metal, $X =$ pnictogen), a class of intermetallic  compounds that crystallize in the tetragonal ZrCuSi$_2$-type structure (space group $P4/nmm$) with a stacking arrangement of Ce$X$-$T$-Ce$X$-$X$ layers. Great attention has been given to the antimonide ($X =$ Sb) members due to the presence of anomalous ferromagnetism in CeAgSb$_{2}$ and, more recently, due to the report of field-induced quantum criticality in CeAuSb$_{2}$ \cite{Onuki,Balicas}. Although the investigation of the antimonides is abundant, fewer reports can be found on the bismuthide ($X =$ Bi) members. Recent studies of high-quality CeCuBi$_{2}$ and CeAuBi$_2$ revealed AFM ordering temperatures of $ T_N=$16~K and 12~K, respectively.
Both materials show a moderate mass enhancement from $T$-linear  electronic specific heat ($\gamma$ on the order of 100 mJ$\cdot$mol$^{-1}$~K$^{-2}$) and exhibit a single spin-flop transition with field applied along the $c$-axis \cite{Adriano2015,PhysRevB.90.235120}.

\begin{figure}[!hb]
\begin{center}
\includegraphics[width=.85\columnwidth]{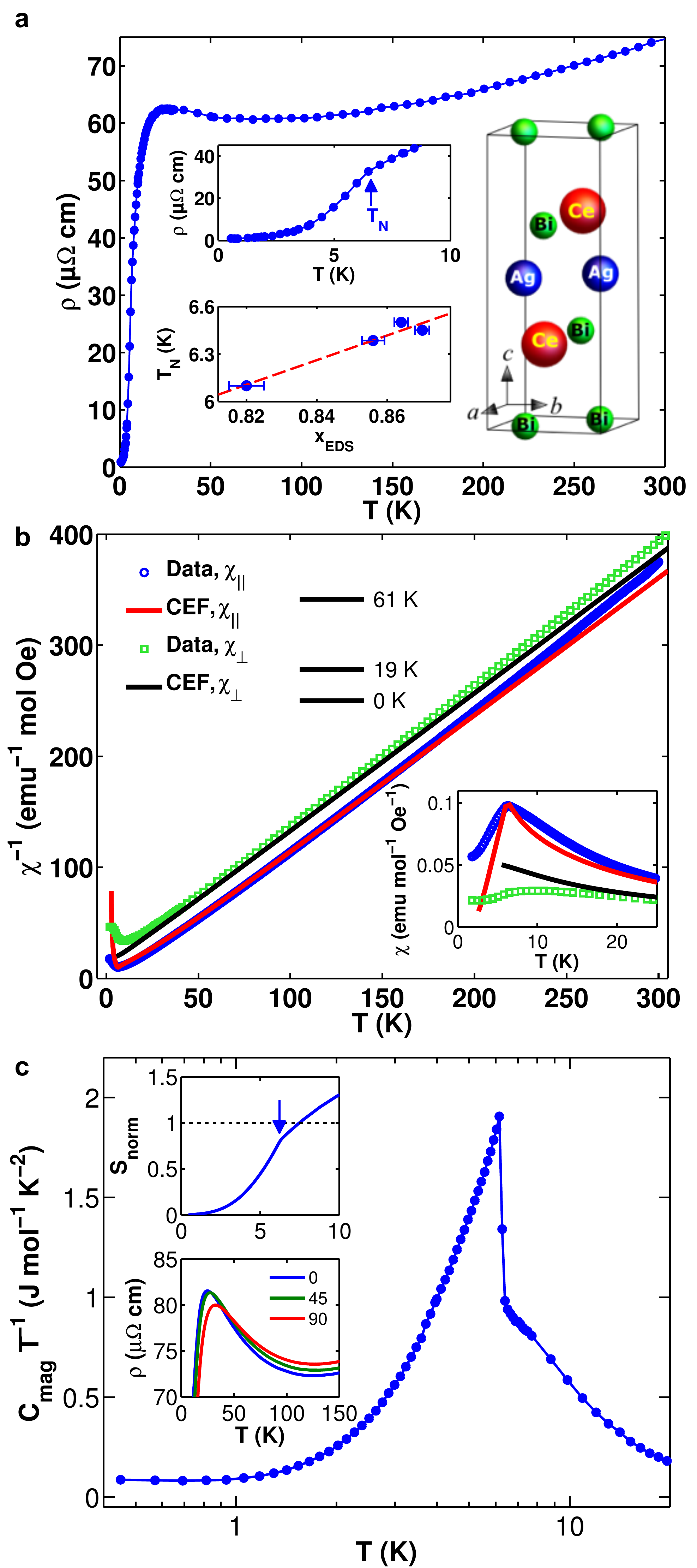}
\vspace{-0.7cm}
\end{center}
\caption{
(a) Temperature dependence of in-plane resistivity, $\rho(T)$, in zero applied field. Top inset shows the low-$T$ region where a kink is observed at $T_{N}$. Bottom inset shows $T_N$ as a function of Ag occupancy measured by EDS. (b)~Inverse susceptibility data for fields parallel ($\chi{}_{\parallel{}}$) and perpendicular ($\chi{}_{\perp}$) to the $c$-axis. Inset shows low temperature $\chi{}(T)$ data. The solid lines are fits to the data using the model described in the text. (c)~Magnetic contribution to specific heat in zero field. Top inset shows the integration of specific heat. $S_{norm}$ is defined as $S/(R\ln{2})$. Bottom inset shows $\rho(T)$ in three different magnetic fields (in kOe).
}
\label{fig:RandChi}
\end{figure}

CeAgBi$_2$, a third member of the isovalent series,  has been reported previously to order antiferromagnetically ($T_N = 6.1$~K) at zero magnetic field and to undergo three field-induced magnetic transitions at 2~K ($H_{c} = 35, 50, 83$~kOe) \cite{Thamizhavel2003,Petrovic2003}. However, a detailed analysis of the underlying interactions and a deep understanding of the phase diagram at low temperatures is still missing.

Single crystals of CeAgBi$_2$ were grown from Bi-flux with starting composition Ce:Ag:Bi=1:$x_{nominal}$:8 ($1 \le x_{nominal} \le 3$).
The mixture was placed in an alumina crucible and sealed in a quartz tube under vacuum.
The sealed tube was heated up to $1050 ^{\circ}\mathrm{C}$ for 8 h and then cooled down at $10 ^{\circ}\mathrm{C/h}$.
The excess of Bi flux was removed by centrifugation after 24 h of annealing at $500 ^{\circ}\mathrm{C}$.
Single crystals with dimensions $\sim 3 \times 3 \times 0.5$ mm$^{3}$ were ground and their crystal structure was checked by X-ray powder diffraction experiments using Cu $K\alpha$ radiation at room temperature.
Several single crystals from different batches were also submitted to elemental analysis using a commercial Energy Dispersive Spectroscopy (EDS) microprobe coupled to a FEG SEM microscope.
From the EDS analysis, we have extracted the actual $x_{Ag}$ concentration.
The precision of the analysis was calculated by $\sigma$/$\sqrt{N}$, where $\sigma$ is the standard deviation of the measurements, and $N$ is the number of points analyzed.

Magnetization measurements were performed using a commercial superconducting quantum interference device (SQUID) down to 1.8 K.
Below 1.8 K, cantilever-based torque magnetometry was used to measure the magnetization in a dilution refrigerator.
Electrical resistivity measurements were made using a low-frequency ac resistance bridge and a four-point configuration.
Hall data was obtained by measuring in both positive and negative applied fields and taking the difference ($\left[R(H+) - R(H-)\right]/2$).
Due to hysteresis in the transitions data was paired for subtraction based on whether the magnitude of the field was increasing or decreasing.
All measurements reported here, except for some of the magnetic susceptibility measurements, were made with the applied field parallel to the c-axis.
For resistivity measurements, the current was applied in the ab-plane.

\section{Experimental Results and CEF fits}

Fig.~\ref{fig:RandChi}a shows the temperature dependence of in-plane resistivity, $\rho(T)$,  down to 0.5 K at zero magnetic field.
At high temperatures ($T \gtrsim 200$~K),  $\rho(T)$ shows metallic behavior, decreasing linearly with decreasing temperature. However, further decrease in temperature reveals a resistivity minimum, followed by a logarithmic increase due to incoherent Kondo scattering. Below $\sim 25$~K, $\rho(T)$ drops abruptly, suggesting that this is the energy scale of either CEF depopulation or Kondo coherence. We will discuss these possibilities below. The kink in resistivity observed at $T_{N} = 6.4$~K, shown in the top inset of Fig.~1a, indicates the transition to the AFM phase.
This represents a 0.3~K higher ordering temperature compared to previous studies of CeAgBi$_2$~\cite{Thamizhavel2003,Petrovic2003}. 
Although this increase in $T_{N}$ is somewhat small, EDS measurements reveal that it is caused by a substantial decrease both in the number of vacancies and in the inhomogeneity at the Ag site.
The bottom inset of Fig.~1a shows the linear dependence of $T_{N}$  on the occupation at the Ag site, $x_{\rm EDS}$.
For the most deficient samples,  with Ag occupation of $82(4)$\%, the transition temperature matches previous studies, $T^{\text{def}}_{N} = 6.1$~K.
For the best samples obtained to date,  the Ag occupancy reaches $87(2)$\%, confirming the trend that transition-metal deficiency is an intrinsic feature of this family of compounds.
We note that less deficient samples are also  accompanied by higher resistance ratios and lower residual resistivity.

Like other Ce-based bismuthides, CeAgBi$_2$ also exhibits magnetic anisotropy.
Fig.~\ref{fig:RandChi}b shows the temperature dependence of the inverse magnetic susceptibility, $1/\chi(T)$, when a magnetic field of $H$ = 1 kOe is applied parallel ($\chi_{\parallel}$) and perpendicular ($\chi_{\perp}$) to the crystallographic $c$-axis. The inset of Fig.~\ref{fig:RandChi}b presents the low-temperature $\chi(T)$ data in which a sharp peak is observed at  $T_{N}$.
The ratio $\chi_{||}/\chi_{\perp} \approx 3.5$ at $T_{N}$ is mainly determined by the tetragonal CEF splitting and reflects the low-$T$ Ce$^{3+}$ single ion anisotropy. This ratio is smaller than what is found in other bismuthides, suggesting a smaller CEF splitting between the ground state and the first excited state as well as a less anisotropic ground state. This scenario will be confirmed below.

At high temperatures ($T > 150$~K), $\chi(T)$ is well-described by a Curie-Weiss (CW) law plus a $T$-independent Pauli term, $\chi(T) = \chi_{0} + C/(T-\theta_{CW})$.
We obtain an effective moment of $\mu_{eff}\approx 2.5(1)\mu_{B}$ for both directions and also for the polycrystalline average, in agreement with the theoretical value of $\mu_{eff}\approx 2.54 \mu_{B}$ for Ce$^{3+}$ free ions.
On the other hand, the $\theta_{CW}$ values are anisotropic, with $\theta_{||}=5.7$~K and $\theta_{\perp}=-8.5$~K.
For the polycrystalline averaged data (not shown), we obtain $\theta_{p} = -4$~K, consistent with AFM order at $\sim$ 6~K.

  \begin{table*}[t]
\begin{tabular*}{0.85\textwidth}{@{\extracolsep{\fill}} cccccccc}
\hline
\multicolumn{8}{c}{CEF parameters} \\
\cline{1-8}\\
& $B_{2}^{0}$ &$B_{4}^{0}$ & $B_{4}^{4}$ & & $z_{FM}J_{FM}$ & & $z_{AFM}J_{AFM}$\\

& -1.78~K& 0.168~K &0.71e-3~K& &-0.89~K& &1.35~K\\
\hline
\multicolumn{8}{c}{Energy levels and wave functions} \\
\cline{1-8}\\
$E (K)$ & & $|-5/2\rangle$ & $|-3/2\rangle$ & $|-1/2\rangle$ & $|+1/2\rangle$ & $|+3/2\rangle$ & $|+5/2\rangle$\\
0 & & 0 & -1 & 0 & 0 & 0 & 0\\
0 & & 0 & 0 & 0 & 0 & 1 & 0\\
19 & & 1 & 0 & 0 & 0 & 0 & 0\\
19 & & 0 & 0 & 0 & 0 & 0 & -1\\
61 & & 0 & 0 & -1 & 0 & 0 & 0\\
61 & & 0 & 0 & 0 & -1 & 0 & 0\\
\end{tabular*}
\caption{
CEF parameters, energy levels, and wave functions of CeAgBi$_{2}$ single crystals obtained from the best fits of the magnetic susceptibility data to the model described in the text.
}
\label{tab:1}
\end{table*}

We now further explore the role of anisotropic interactions and CEF effects in the magnetic properties of CeAgBi$_2$. To this end, we analyze the experimental data using a mean field model including two mean anisotropic interactions ($z_{\mathrm{AFM}}J_{\mathrm{AFM}}$ and $z_{\mathrm{FM}}J_{\mathrm{FM}}$) between nearest-neighbors, which do not contain directional information.
We also take into account the tetragonal CEF Hamiltonian $H_{CEF}=B_{2}^{0}O_{2}^{0} + B_{4}^{0}O_{4}^{0} + B_{4}^{4}O_{4}^{4}$, where $z$ is the number of nearest neighbors, $B_{i}^{n}$ are  the CEF parameters, and $O_{i}^{n}$ are the Stevens equivalent operators obtained from the angular momentum operators.
A more detailed description of the model can be found in Ref. \cite{Pagliuso_JAP2006}.

This model was used to simultaneously fit $\chi(T)$, $M(H)$ and $C_{mag}(T)/T$ data in the entire range of temperature.
The best fit which reproduces the anisotropic susceptibility is shown by solid lines in Fig.~\ref{fig:RandChi}b and the extracted parameters for the CEF scheme and the exchange interactions are given in Table~\ref{tab:1}.
We find a $|J=5/2,J_z = \pm3/2\rangle$~Kramers doublet ground state, separated from other excited doublets by 19~K and 60~K.
We also find, in fact, (i) the CEF splitting is much smaller as compared to other bismuthides and (ii) the ground state is mainly $|J=5/2,J_z = \pm3/2\rangle$ instead of the dominantly $|J=5/2,J_z = \pm5/2\rangle$ found in CeCuBi$_{2}$ and CeAuBi$_{2}$.
Further, the dominant CEF parameter $B_{2}^{0}$ obtained from the fits is similar to the value obtained using the high-temperature expansion of $\chi$: $B_{2}^{0}=10(\theta_{\perp}-\theta_{||})/[3(2J-1)(2J+3)]=-1.48$~K.\cite{Avila2003}
This result suggests that the effects of anisotropic interactions at high temperatures are smaller in CeAgBi$_2$ than in CeCuBi$_2$ and CeAuBi$_2$.
We also note that although the magnetic anisotropy and $T_{N}$ along the $c$-axis are well reproduced by our simple model, the single ion CEF effect is not able to capture all field-induced transition in $M(H)$ data at 1.8~K.

Fig.~\ref{fig:RandChi}c shows the temperature dependence of the specific heat of CeAgBi$_2$ in zero field. To determine the magnetic contribution to specific heat, the data obtained from non-magnetic reference compound LaAgBi$_2$ was subtracted from the result. There is a single peak corresponding to T$_N$. We note that this sample is from an earlier batch, so it has higher silver vacancies and a lower T$_N$.
The top inset of Fig.~1c shows the integration of specific heat over temperature, i.e., the recovered magnetic entropy.
At T$_N$, the recovered entropy is only about 80\% the ground-state doublet (R$\ln{2}$).
We note, however, that there is a broad feature centered around 8~K, which is consistent with a Schottky anomaly generated by the CEF splitting of 19~K.
This result indicates that the first excited CEF state is already partially occupied at T$_N$, giving an additional entropy to the expected R$\ln{2}$.
Therefore, the reduction of entropy is larger than 20\%, although a precise calculation cannot be evaluated.
It is unlikely that a reduction of entropy of more than 20\% can be fully accounted for by classical magnetic fluctuations.
For example, numerical calculations of the Ising model on a simple cubic lattice find an entropy reduction of just under 20\% at T$_N$, with other three dimensional geometries showing an even smaller reduction.\cite{DeJongh1978}
We thus attribute this reduction to both magnetic frustration and hybridization between Ce $4f$-electrons and conduction electrons.
This conclusion is motivated by the enhanced effect masses found in dHvA experiments\cite{Thamizhavel2003} and the magnetic frustration that often arises from competing exchange interactions.

Due to the similar energy scales of the first excited CEF state ($19$~K) and the peak in resistance ($24$~K), we performed high-temperature resistance measurements in several different fields.
The results are presented in inset of Fig.~1c.
If the peak near 24~K were due to crystal field depopulation, then increasing the field should lead to an increase of the energy difference between
the lowest Zeeman split state of the ground state and the upper Zeeman split state of the first excited CEF state.
This in turn would lead to an increase of the temperature of the resistance peak as the field is increased, which is in fact observed experimentally.
We note that, in heavy fermion compounds where the energy difference between Kondo coherence and $\Delta_{\mathrm{CEF}}$ is small, the onset of Kondo coherence ($T^{*}$) usually occurs at temperatures of the order of $\Delta_{\mathrm{CEF}}/2$.
This suggests that $T^{*} \sim 10$~K in CeAgBi$_{2}$.
However, the experimental lack of two distinct resistance maxima as is typically observed when the energy difference is large\cite{Lassailly1985} leads to an ambiguity as to whether the resistance peak is due to CEF splitting of Kondo coherence.

\begin{figure}[!hb]
\begin{center}
\includegraphics[width=0.95\columnwidth]{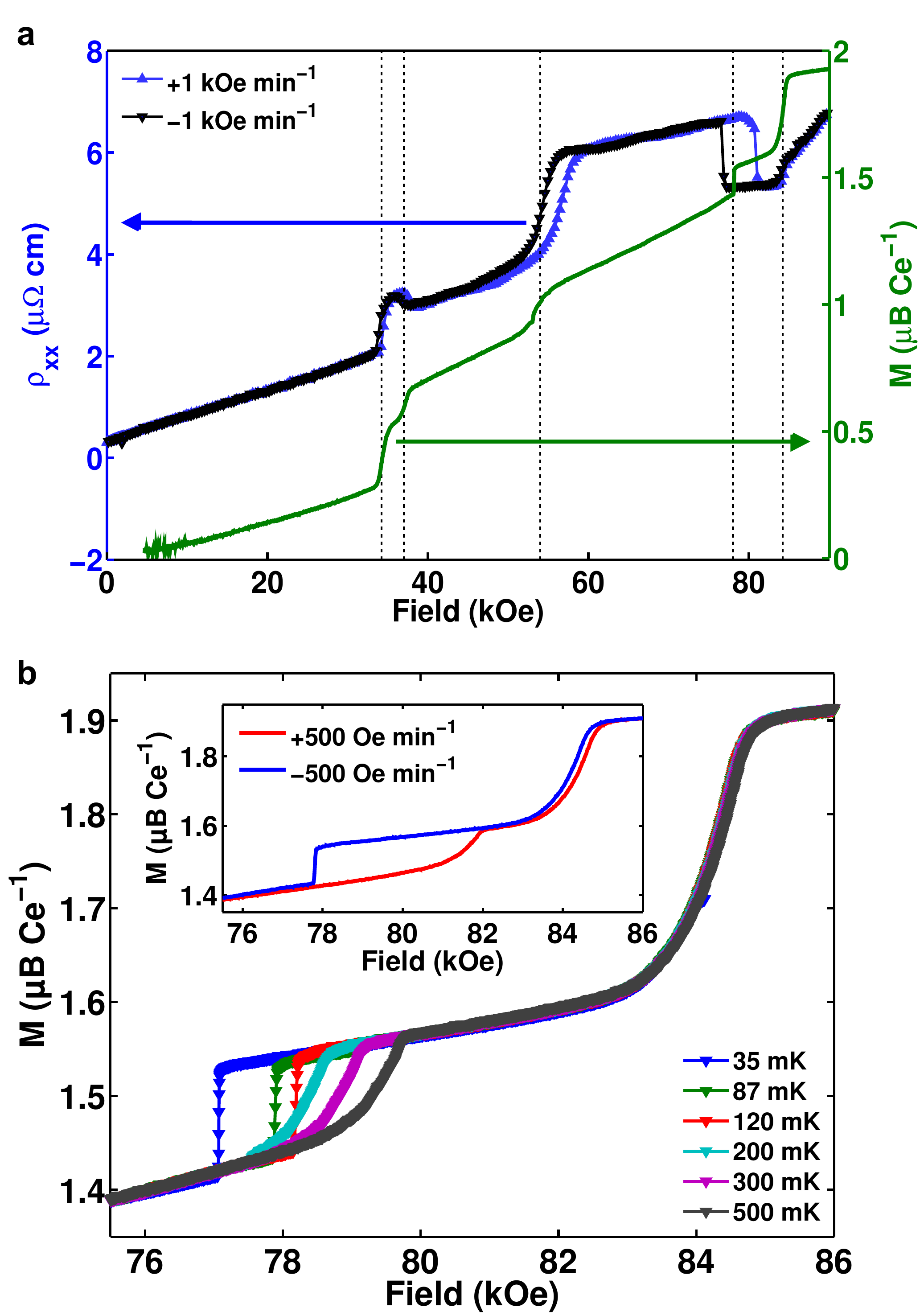}
\vspace{-0.7cm}
\end{center}
\caption{
Field-tuned magnetic and transport properties of CeAgBi$_2$.
(a)~$\rho_{xx}$ (blue) and magnetization (green) versus field at 100~mK. $\rho_{xx}$ data is for both increasing and decreasing fields, showing hysteresis for transitions centered near 54 and 78~kOe.
(b)~Magnetization versus field for decreasing field-sweep at temperatures 500~mK and lower. Sweep rate was 500~Oe/min. Inset shows hysteresis in magnetization depending on field sweep direction.
}
\label{fig:MagandMR}
\end{figure}

We now turn to the analysis of the low-temperature phase diagram as a function of an applied magnetic field along the $c$-axis. Fig.~\ref{fig:MagandMR}a shows magnetization and (transverse) magnetoresistance (MR) of CeAgBi$_{2}$ at 100~mK. As the field is swept from $0$ to $90$~kOe, we  find a sequence of field-induced metamagnetic transitions near $H=34, 37, 54, 78,$ and $84$~kOe, indicating a complex  phase diagram for $T<T_N$. Above the $84$~kOe transition, the magnetization saturates just below $1.9$ $\mu_{B}$/Ce, which is slightly lower than the saturation value of $2.1$ $\mu_{B}$/Ce previously reported, and may be due to a non-linearity between the deflection of the cantilever and the corresponding change in capacitance.
Such large saturated value is somewhat surprising since the largest possible $c$-axis magnetization of a $|\pm3/2\rangle$  CEF ground state is $(3/2)g_L.\mu_B$, i.e., 1.29  $\mu_{B}$/Ce. Hence, our results indicate that the Zeeman effect induces a change in the CEF ground state from $|\pm3/2\rangle$ to $|\pm5/2\rangle$ when  $H \sim 84$~kOe.

Of the observed transitions, the one near 78~kOe appears to be strongly first-order as shown by magnetization in  Fig.~\ref{fig:MagandMR}b. As $T$ is lowered from 500~mK to 35~mK a sharpening of the transition is observed and the transition eventually becomes step-like.
Field sweeps performed around this transition show signs of irreversibility, namely a $\sim4$~kOe wide hysteresis loop,  providing further evidence of first-order behavior.
This in contrast with other Ce-based bismuthides. For instance,  other members of the isovalent series such as CeCuBi$_2$ ($T_{N} =  16$~K) and CeAuBi$_2$ ($T_{N} =  12$~K) show only a single clear spin-flop transition. We note, however, that multiple steps have been observed in previous reports on more deficient CeCuBi$_2$ ($T_{N} =  11$~K) and even CeCuBi$_2$ ($T_{N} =  16$~K) shows hints to a second transition just before the main one.
It is also noteworthy that hysteresis loops show a significant difference ($\sim 2$ kOe) at the transition centered near $54$~kOe as well. The change in magnetization, however, is not nearly as abrupt as found in the 78~kOe transition. It is well-known that transport measurements in real materials are complicated by disorder. Thus, as this transition is clearly not strongly first-order, the slight amount of irreversibility is likely attributed to pinning due to crystallographic defects (e.g., inherent silver deficiency). In fact, a small amount of hysteresis (on the order of a hundred gauss) was also observed in magnetization and transport at the remaining three transitions.

The rich phase structure in CeAgBi$_2$ is likely due to the weak anisotropy combined with anisotropic exchange parameters with opposite signs, which lead to magnetic frustration. Further, the close energy scales of the CEF splitting ($\sim19$~K), Kondo coherence  ($\sim 10$~K) and the AFM ($T_N\sim6.4$~K) generates a complex response of the physical properties to the application of magnetic field.

The MR also reflects the multi-step phase structure, tracking each metamagnetic transition with a sharp step, indicating the presence Kondo coupling between the Ce magnetic  moments and itinerant $p$-electrons from Bi. The overall trend of MR is linear, a feature shared with the (non-magnetic) compounds LaAgBi$_2$ and LaAgSb$_2$, where it is attributed to an underlying Dirac dispersion for Bi/Sb itinerant electrons~\cite{wang2013quasi,myers1999systematic}. A preliminary ab initio calculation~\cite{SBLee2015}  for CeAgBi$_{2}$ indicates that the Bi $p$-electrons also have a Dirac dispersion, suggesting a similar origin for the underlying linear MR. However, we note an alternative explanation for the increase in MR is an enhancement in spin-disorder scattering in AFM materials in an applied field~\cite{Phys.Rev.168.531,yamada1973magnetoresistance}. Note that the MR exhibits a sharp drop and hysteretic behavior near the  $79$~kOe transition, lending further support that it is first-order.

\begin{figure}[!htb]
\begin{center}
\includegraphics[width=0.95\columnwidth]{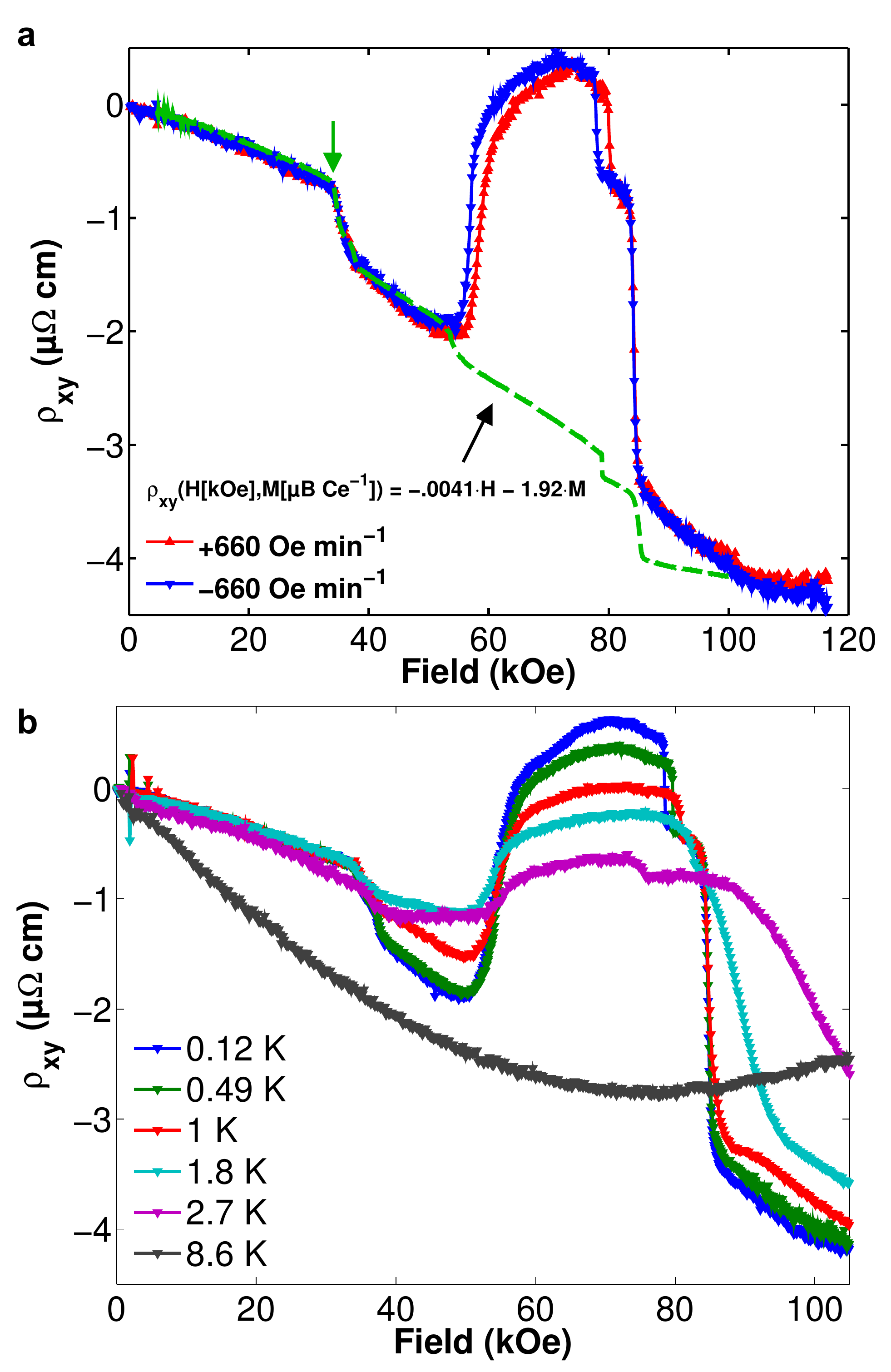}
\vspace{-0.7cm}
\end{center}
\caption{
Hall resistance ($\rho_{xy}$) versus field.
(a) Comparison of sweeping the field up vs. down at 150~mK. The dotted line is a least-squares fit to $\rho_{xy}(H) = R_{H}H+R_{M}M(H)$. The fit was made to the region of zero field up to 37~kOe (green arrow) and extrapolated beyond that point.
(b) In a second sample, comparison at different  $T$ while sweeping the field downward.}
\label{fig:Halldata}
\end{figure}

Motivated by this evidence for coupling of itinerant electrons and local moments, we performed measurements of the Hall resistivity $\rho_{xy}$ to further elucidate the nature of the different phases.
Like the MR, the Hall effect tracks each metamagnetic transition (Fig.~\ref{fig:Halldata}).

The decrease in $\rho_{xy}$  roughly parallels the increase in magnetization over the first two transitions, but then sharply  deviates from this trend as the field is increased across the trio of transitions at 57, 78, and 84~kOe, even changing its sign between $\sim{}60-80$~kOe.
Further insight is afforded by fitting the Hall resistivity to the standard form $\rho_{xy} (H) = R_H H + R_M M(H)$ that includes  Hall effect contributions from both the applied field and the induced magnetization.
Using the data up to $H\lesssim 38$~kOe to obtain the fit parameters, we find that this model fits the data extremely well for low fields $H\lesssim54$~kOe, whereupon the measured $\rho_{xy}$ strongly deviates from the expectation based on the model.
While there is still some discrepancy between the fit and the data after the last transition  ($H\gtrsim 84$~kOe), this is relatively small compared to the much larger deviation between fit and data in the regime $54$~kOe~$\lesssim H\lesssim 84$~kOe.
As the temperature is raised, the magnitude of the anomaly in $\rho_{xy}$ near 54~kOe is decreased, as shown in Fig.~\ref{fig:Halldata}b.
At present, we do not have a complete understanding of the origin of this Hall effect anomaly, but we comment on some possible scenarios below.
We note that while there is an additional deviation from linear behavior of $\rho_{xy}$ above $\sim110$~kOe, a previous study on lower-quality samples using pulsed fields up to $400$~kOe~\cite{Thamizhavel2003} appears to rule out an additional transition.

\section{Discussion}

\begin{figure}[!ht]
\begin{center}
\includegraphics[width=0.9\columnwidth]{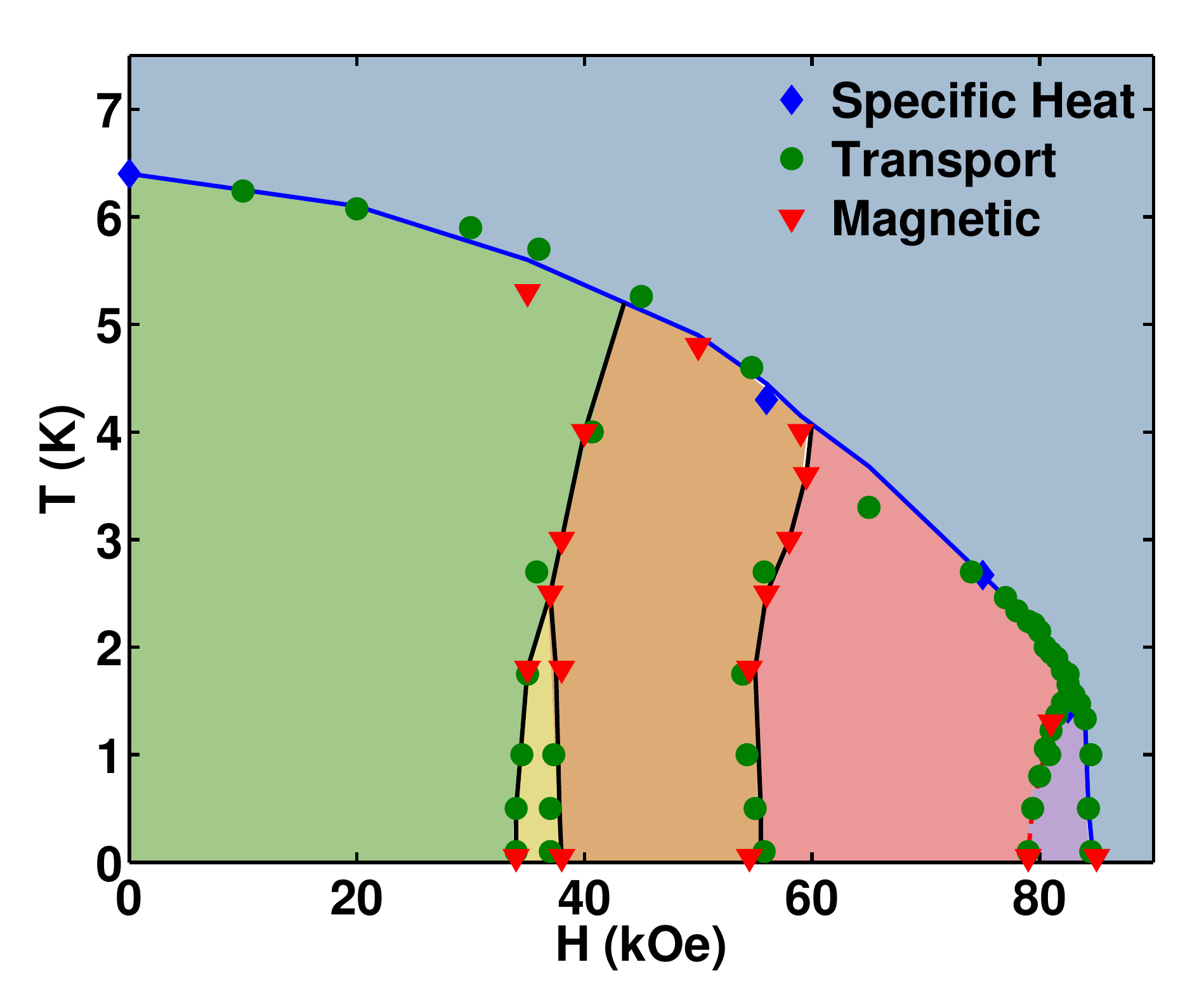}
\vspace{-0.7cm}
\end{center}
\caption{
(a) Low temperature H-T phase diagram from CeAgBi2. Phase boundary was determined by resistivity, magnetization, and heat capacity measurements.}
\label{fig:phase_diag}
\end{figure}

Our results can be summarized in the form of a phase diagram of CeAgBi$_2$ with temperature and applied field, as shown in Fig.~\ref{fig:phase_diag}.
The second $T=0$ phase (for 34~kOe~$<H<$~37~kOe) is not seen above the temperature $T\gtrsim$ 2.5~K, illustrating the fragility of this phase to thermal fluctuations.
Note the strong agreement between the probes of specific heat, transport, and magnetic structure when the corresponding measurements overlap.
Of the sequence of transitions between the different low-temperature phases, the penultimate transition near 79~kOe appears to be strongly first-order, whereas the remaining transitions appear continuous.
As noted, the Hall effect is well-described by a simple model including effects of both the applied field and induced magnetization very well below $54$~kOe and reasonably well above $84$~kOe, but exhibits a strong deviation from this simple behavior in the intermediate field regime.

Heavy fermion compounds CeAuSb$_2$ and YbAgGe both have some commonalities in the low-temperature phase diagram.
For instance, CeAuSb$_2$ orders AF and has at least one magnetic transition before the suppression of AF order to 0~K at 54~kOe.\cite{Balicas}
Recent work suggests that the first transition might bifurcate into two at lower temperatures, similar to CeAgBi$_2$.\cite{Lorenzer2013}
YbAgGe also has a similarly complicated phase diagram as a function of field which contains a region partially bordered by a first order transition \cite{Budko2003a}.  Interestingly,  in this region YbAgGe exhibits anomalous behavior attributed to the influence of a quantum bicritical point \cite{BCP}.
Unlike in YbAgGe, however, the set of transitions is only observed in CeAgBi$_2$ for $H||c$.
Further, the large anomalies in the Hall resistivity appear to be a feature unique to CeAgBi$_2$.

We now comment on potential explanations for the physics at hand.
First, we attribute the sequence of metamagnetic transitions to the frustration from competing anisotropic exchange interactions, as evinced by the susceptibility data. As in many HFs, the Hall data is more challenging to describe \cite{NairHFHall}.
Its evident sensitivity to the magnetization suggests a significant contribution from the induced magnetization at low fields, yet a simple phenomenological model incorporating this contribution seems unable to describe the behavior at intermediate fields.
Here we note that quantum oscillation measurements reveal electron effective masses in the range of 5 to 7 m$_e$, suggesting heavy fermion physics plays a role in transport~\cite{Thamizhavel2003}; this could perhaps explain the anomaly, e.g. via a Fermi surface reconstruction. To explore this scenario, one would have to investigate, for instance, band-structure calculations in the presence of a magnetic field. We anticipate that future theoretical analysis, incorporating details of the conduction electron band structure, as well as a careful treatment of their coupling to local moments, will shed more light on these issues~\cite{SBLee2015}.


An alternative scenario involves the `geometrical Hall effect' from a non-collinear spin texture; while more exotic, such physics is now quite well-established as a consequence of Dzyaloshinskii-Moriya interactions in ferromagnets and has been also proposed to occur in anisotropic exchange AFMs{, and recently also in heavy-fermion metals~\cite{DingAHE_HF2015}.} In this case, an anomalous Hall effect  could be observed as a result of a non-vanishing Berry-phase curvature \cite{AHE_AFM}. To this end, neutron diffraction/scattering experiments would be valuable to unveil, for instance, the field regions where non-collinear spin structures are realized.

\section{Spin model and metamagnetic transitions}

As mentioned above, our previous model was unable to capture all field-induced magnetic transitions. Finally, in the present section we introduce a spin model for the Ce local moments to discuss their field-induced response.
Although more experimental investigation is clearly essential in order to clarify what type of AFM order is stabilized in CeAgBi$_2$ in each field region, here we sketch one possible scenario built on existing experimental data on the closely-related compound CeCuBi$_2$, where the zero-field magnetic structure has already been well-established experimentally \cite{PhysRevB.90.235120}.
We defer a detailed analysis of the spin model to future work, and here only discuss a simple mean-field picture of the magnetic order.

\begin{figure}[!hb]
\begin{center}
\includegraphics[width=0.35\columnwidth]{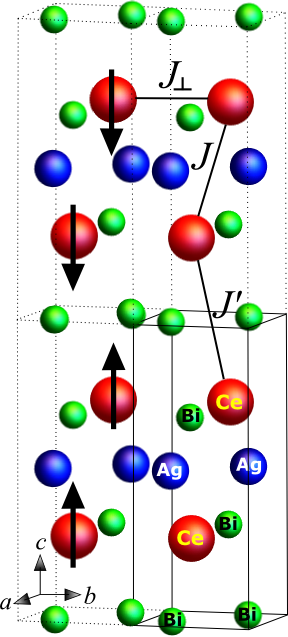}
\vspace{-0.7cm}
\end{center}
\caption{Lattice structure of CeAgBi$_2$ : Red, blue and green spheres represent Ce, Ag and Bi ions respectively and solid line shows one unit-cell. ``up-up-down-down" like ordering of Ce moments is also shown along $c$ axis, which is measured to be the zero field magnetic structire in CeCuBi$_2$. $J$, $J'$ and $J_\perp$ indicates three symmetry distinct spin exchange interactions between Ce magnetic moments.}
\label{suppfig:1}
\end{figure}

For CeCuBi$_2$,  X-ray magnetic diffraction reveals that the Ce local moments order antiferromagnetically with an ordering wave vector $(0,0,0.5)$~\cite{PhysRevB.90.235120}. This corresponds to  ``up-up-down-down"  magnetic ordering along the crystallographic $c$-axis and ferromagnetic (FM) ordering across the basal $a$-$b$ plane.
(See Fig.~\ref{suppfig:1}).
The magnetic susceptibility data shows a large anisotropy consistent with a strong easy-axis along the $c$-axis.
The application of a $c$-axis magnetic field results in a single spin-flop transition at a field  of $\sim 55$kOe; for fields in the $ab$ plane, no such transition was observed.
 Within a classical approximation, a minimal spin model that may be inferred from these experimental results incorporates a single-ion anisotropy term and dominant spin exchanges between neighboring sites:
\begin{equation}
{\mathcal H} = \sum_{ ij } J_{ij} {\bs S}_{i} \cdot  {\bs S}_{j} -  \sum_{i}  {\bs h} \cdot {\bs S}_{i} - \Delta \sum_{i} ({S_i}^z )^2.
\label{suppeq:1}
\end{equation}
Here,  ${\bs h}$ is the (applied) magnetic field and $\Delta>0$ is the magnitude of an easy-axis single-ion anisotropy along the $c$-axis. $J_{ij}$ represents competing spin exchanges between neighboring sites;
FM and AFM (or AFM and FM) spin exchanges for two symmetry distinct neighbors along $c$ axis are denoted  $J $ and $J' $, respectively, and a FM spin interaction in the $a$-$b$ plane is denoted as $J_\perp$, as seen in Fig.~\ref{suppfig:1}.
The competition between alternating ferro- and antiferromagnetic spin exchanges along the $c$-axis stabilizes the ``up-up-down-down"  magnetic structure along $c$-axis, with a spin polarized state FM on each $a$-$b$ plane.
In the limit $\Delta \gg J_{ij}$, the moments behave like $c$-axis Ising spins, and the presence of a $c$-axis field leads to a direct magnetic transition from AF ordering to a fully-spin-polarized, paramagnetic state.
However, a smaller single ion anisotropy $\Delta \lesssim J_{ij}$ admits a spin-flop transition at finite $c$-axis field, resulting in AF spin canting.
This is consistent with the experimental data. Given that the minimal picture of competing spin exchanges and single ion anisotropy provide a  reasonable explanation of magnetic ordering and field effect of CeCuBi$_2$ material, we now turn to a similar analysis of CeAgBi$_2$.

Magnetic susceptibility measurements on CeAgBi$_2$ again show the presence of magnetic anisotropy but smaller than in CeCuBi$_2$, which indicates weaker Ising anisotropy of the localized moments.
Furthermore, when Ag substitutes for Cu on the the transition metal site, spin superexchange between Ce moments occurs via the $4d$ or $5s$ Ag orbitals, as compared to the $3d$ or $4s$ electrons of Cu, thereby changing the effective superexchange couplings $J, J', J_\perp$.
Moreover, the slight difference in lattice parameters and bond angles may also lead to changes in the RKKY interactions induced by coupling to the itinerant electrons of Bi. In general, this confluence of competing effects may stabilize very complicated magnetic ordering.
We also note that further-neighbor interactions besides $J$, $J'$ and $J_\perp$ generally induce incommensurate {\it spiral} ordering, affording an even richer set of possibilities. For the purposes of a preliminary analysis, we focus on our minimal spin model keeping  just $J$, $J'$ and $J_\perp$, and describing the possible ordered states as a function of the applied field. Notably, even this simple model exhibits a plethora of magnetic orders separated by a sequence of metamagnetic transitions.

A convenient approach to understand the magnetic order is to separately consider every two layers of Ce sites, corresponding to a single unit cell lattice spacing along the $c$ axis.
These two layers of Ce sites then form a buckled square lattice, where the nearest and next-nearest neighbor spin-spin couplings are $J$ and $J_\perp$ respectively; we will refer to this as a `bilayer'.
Reasoning in analogy with the well-studied $J_1-J_2$ square-lattice Heisenberg model, we anticipate that FM, N\'eel and stripy phases can be stabilized in each bilayer depending on the  relative signs and magnitudes of $J$ and $J_\perp$.
We may then use these orders as a building block to construct the full three dimensional magnetic order by coupling the Ce bilayers via an interlayer coupling $J'$.
We now briefly summarize how competing interactions and magnetic anisotropy can lead to several metamagnetic transitions within this approach, even if we restrict ourselves to working within a classical spin approximation.
In order to have several metamagnetic phases, we require that $J$, $J'$, $J_\perp$ and $\Delta$ are all of the same order; this is also consistent with expectations based on experiments and a study of lattice parameters.
To describe the ordered states stabilized with this parameter regime, it is convenient to work with a magnetic unit cell 8 times larger than the original unit cell (i.e., containing 16 Ce sites), and optimize the classical energy of this unit cell for a given set of parameters $J_{ij}$, $\Delta$ and $h$, we minimize the energy of Eq.\eqref{suppeq:1}.

\begin{figure}[!hb]
\begin{center}
\includegraphics[width=0.95\columnwidth]{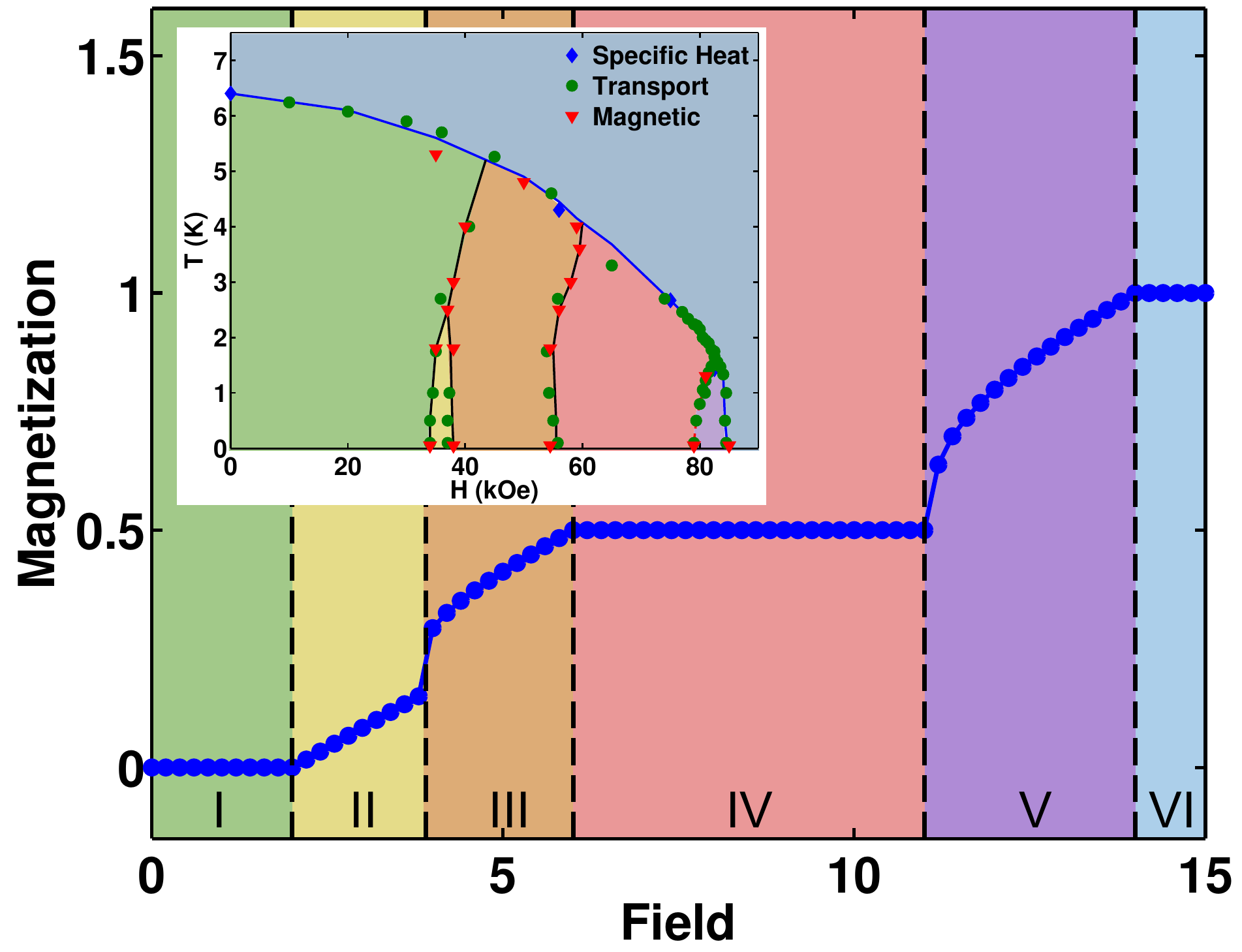}
\vspace{-0.7cm}
\end{center}
\caption{Plot of magnetization $m$ {\it vs} field $h$ applied along $c$ axis. In the presence of competing spin exchange interactions and anisotropy, there exist several magnetic phases with applying fields. (In particular, we take $J = -1$, $J'=2$, $J_\perp= 1.5$ and $\Delta =1$ in Eq.\eqref{suppeq:1})  Fig.~\ref{suppfig:3} describes different magnetic ordering patterns in every phase, I through VI.
Inset shows experimental phase diagram from Fig.~4.
}
\label{suppfig:2}
\end{figure}

\begin{figure*}[!hb]
\includegraphics[width=2\columnwidth]{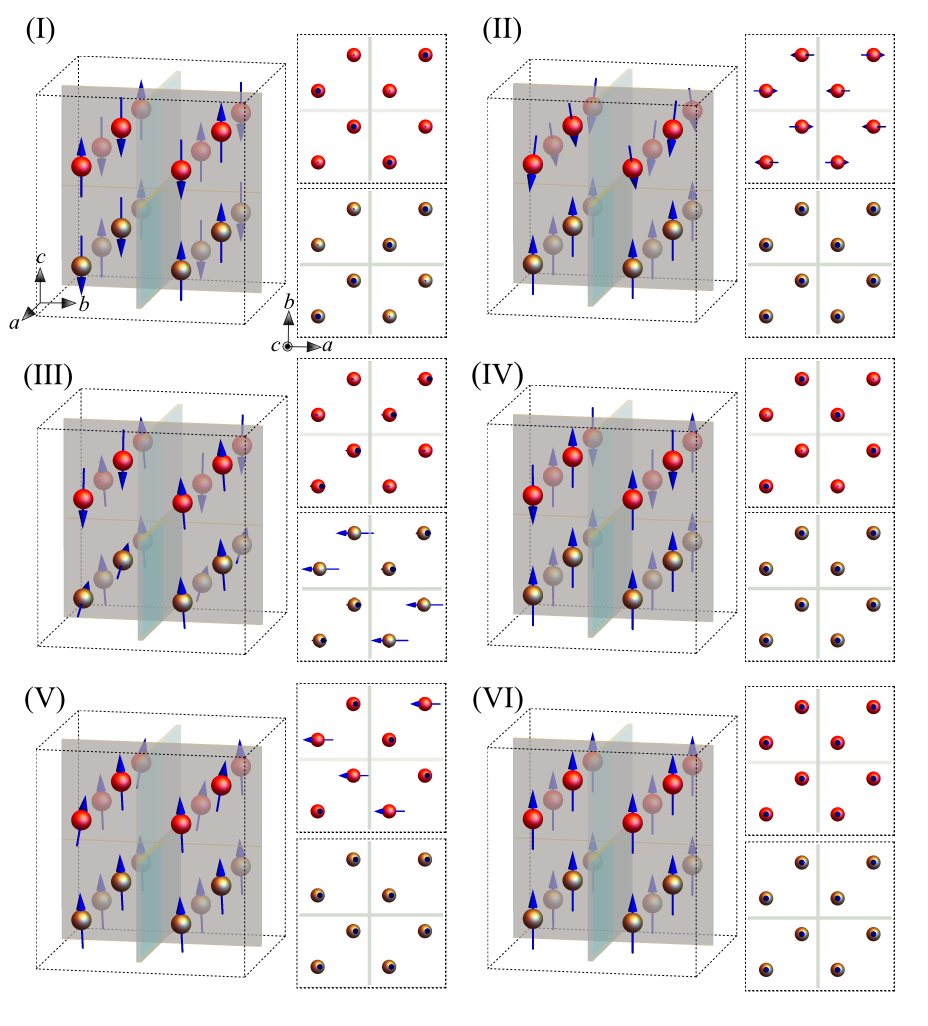}
\vspace{-0.7cm}
\caption{ {\bf Magnetic ordering patterns for the phases (I) through (VI) in Fig.~\ref{suppfig:2}.} For a clear description of ordering patterns, three figures are shown for every phase: the  figure at left shows the ordering pattern in the 16-site magnetic unit cell, while the two figures on the right are  top-down views of upper and lower Ce bilayers, respectively. (I) Spins in every bilayer form a stripy phase on a buckled square lattice. (II) Spins in the lower bilayer are polarized along the field direction while spins in the upper bilayer cant in the field direction. A small canting induces the finite magnetization $m$ as seen in Fig.~\ref{suppfig:3}. (III) Spins in the upper bilayer form a stripy phase and four spins flop in the lower bilayer, resulting in a partial spin-flop transition with a finite magnetization jump at the transition point $h \approx 4$. (IV, Half magnetization plateau region) Spins in the lower bilayer are polarized along the field direction and spins in the upper bilayer form a stripy phase. (V) Spins in the lower bilayer are polarized and four spins in the upper bilayer flop from a stripy phase, resulting in another partial spin-flop transition and magnetization jump near $h\approx 11$.  (VI) All spins are fully polarized along the field direction beyond the saturation field.}
\label{suppfig:3}
\end{figure*}

Fig.~\ref{suppfig:2} shows the magnetization $m$ ($m=1$ is the fully polarized classical spin per site) as a function of $c$-axis  magnetic field for specific values of parameters $J = -1$, $J'=2$, $J_\perp= 1.5$ and $\Delta =1$, which are the parameters that best describe the experimental data.
In fact, we find six distinct phases over our field range.
Note that the exact same argument can be applied when $J$ and $J'$ values are interchanged, due to site connectivity.
The magnetic ordering pattern in each of these phases is illustrated in Fig.~\ref{suppfig:3}.

At a zero field, each buckled square lattice bilayer forms a stripy phase, as $|J| < |J_\perp |$ and the stripy phase is robust under a small and finite magnetic field.
With the chosen parameters, this phase is degenerate in energy with the ``up-up-down-down" like ordering (discussed for the case of CeCuBi$_2$ and illustrated in Fig.~\ref{suppfig:1}).
Further increase of magnetic field, however, induces a transition into a phase where spins in the lower two layers are polarized along the field direction and spins in the upper two layers are anti-aligned to the field but have a finite canting.
The canted moments induce a finite magnetization, preventing perfect cancellation of magnetic moments between the upper and lower bilayers. At $h \approx 4$, a partial spin-flop transition occurs, yielding a finite magnetization jump.
In this phase, spins in the upper two layers almost form a stripy phase, similar to the phase (I). In the lower two layers, however, all spins attempt to align along the field direction but four spins form a different canting angle with respect to the other four.
By smoothly changing those canting angles, this phase reaches a half-magnetization plateau region, where spins in the upper bilayer form a stripy phase and spins in the lower bilayer are fully polarized.
With stronger magnetic fields, there is another partial spin-flop transition exhibiting a finite jump in the magnetization.
In phase (V), all 16 spins attempt to be aligned along the field direction but four of them have a different canting angle until their canting angle becomes zero and a fully polarized spin state (VI) is realized at a saturation field $H \approx 14$.

Based on the above analysis, we conclude that competing exchange interactions and anisotropy can in principle lead to several metamagnetic transitions with field.
We exemplified this for a specific parameter regime in Eq.\eqref{suppeq:1} and found six distinct phases that may emerge taking into account the most dominant exchange interactions.
Future neutron scattering studies will be valuable to confirm such scenario, and lay a firmer foundation for theoretical studies.
More detailed theoretical and experimental analyses will be discussed in Ref.~\onlinecite{SBLee2015}, which will also discuss scenarios beyond this minimal model and attempt to connect the magnetic ordering with transport.

\section{Conclusions}

In conclusion, we have shown that high-quality CeAgBi$_{2}$ single crystals  (${T_{N} = 6.4}$~K)  present a rich field-tuned phase diagram with five metamagnetic transitions at $40$~mK. In contrast to other Ce$TX_{2}$ members, a strongly first-order transition is observed at $\sim 79$~kOe in the vicinity of the transition to the paramagnetic state. Remarkably, we unveil an anomalous Hall contribution for fields $H<54$ kOe which is drastically altered when $H$ is tuned through a trio of transitions at 57, 78, and 84~kOe, suggesting that the Fermi surface is reconstructed in a subset of the metamagnetic transitions. Our results shed light on hidden properties of CeAgBi$_{2}$ and open new avenues for both experimental and theoretical studies on non-collinear magnetic structures, quantum (bi-)criticality, and Fermi surface effects, to name a few.

\begin{acknowledgments}
The authors thank S. Raghu, S. Kivelson and Leon Balents for discussions. This work was supported by NSF grant DMR-1350122. P. F. S. R. acknowledges FAPESP grant 2013/17427-7. J. X. acknowledges Sloan Fellowship grant BR2013-116. S.A.P. acknowledges support from NSF grant DMR-1455366

\end{acknowledgments}


\end{document}